\documentclass[prl,aps,superscriptaddress,longbibliography,groupedaddress,twocolumn]{revtex4-2}
\usepackage[colorlinks=true,citecolor=blue,linkcolor=red,urlcolor=red]{hyperref}
\usepackage[sort&compress]{natbib}
\usepackage{graphicx}
\usepackage{amssymb}
\usepackage{braket}
\usepackage{amsmath}
\usepackage{hyperref}
\usepackage{dsfont}
\usepackage{bm} 
\usepackage{physics}
\usepackage{tabularx}
\usepackage{verbatim}
\usepackage{float}
\usepackage{romannum}
\usepackage{bbold}
\usepackage{mathtools}

\renewcommand\k{{\bf k}}

\newcommand\x{{\bf x}}

\begin{document}
\author{Anirudh Gundhi}
\email{anirudh.gundhi@units.it}
\affiliation{Department of Physics, University of Trieste, Strada Costiera 11, 34151 Trieste, Italy}
\affiliation{Istituto
	Nazionale di Fisica Nucleare, Trieste Section, Via Valerio 2, 34127 Trieste,
	Italy}
\author{Hendrik Ulbricht}
\email{H.Ulbricht@soton.ac.uk}
\affiliation{School of Physics and Astronomy, University of Southampton, Southampton SO17 1BJ,
United Kingdom}
	
\title{Measuring Decoherence Due to Quantum Vacuum Fluctuations}
	
\date{\today}

\begin{abstract}
The interaction of a particle with vacuum fluctuations--which theoretically exist even in the complete absence of matter--can lead to observable irreversible decoherence if it were possible to switch on and off the particle charge suddenly. We compute the leading order decoherence effect for such a scenario and propose an experimental setup for its detection. Such a measurement might provide further insights into the nature of vacuum fluctuations and a novel precision test for the decoherence theory.
\end{abstract}

\maketitle

\textit{Introduction--}From an open quantum system perspective, the zero-point modes of the electromagnetic (EM) field comprise an unavoidable environment. Decoherence due to such an environment has been discussed by several works \cite{Caldeira1991, Kiefer1992,Ford1993, Santos1994,Diosi1995,ELZE1995,BandP,Unruh_Coherence,Mazzitelli2003,Hsiang2006,Baym_Ozawa,Gundhi:2023vjs,Gundhi2024Casimir}, with arguments presented both for and against its possibility. Most recently, starting from the nonrelativistic QED Lagrangian, the off-diagonal elements of the reduced density matrix of an electron interacting with the EM field vacuum were calculated in \cite{Gundhi:2023vjs,Gundhi2024Casimir}. There, contrary to \cite{Caldeira1991,Ford1993,Baym_Ozawa,Santos1994,BandP,Mazzitelli2003,Hsiang2006}, and in agreement with \cite{Diosi1995, Unruh_Coherence}, it was shown that the decay of the matrix elements due to vacuum fluctuations (VFs) does not correspond to a genuine loss of coherence that could be observed in a typical interference experiment. 

Physically, VFs manifest themselves as a cloud of virtual photons (electron dressing) that moves with the bare electron \cite{Diosi1995,Kiefer2003}. The quantum vacuum is therefore correlated to the electron's  position but not to its history or trajectory. In a typical double slit experiment, as the electron reaches the detector screen, such an environment would \textit{forget} which hole the electron passed through and cause no observable decoherence \cite{Gundhi2024Casimir}. This can change, however, if the interaction of a particle with the EM field is suddenly switched on and off during the interference experiment, not allowing VFs to \textit{adjust} completely to the system position as they typically do.  We emphasize that  switching the interaction on and off with the environment of VFs is nontrivial to realize, because a charged particle can never \textit{escape} the vacuum itself.

\begin{figure}[!ht]
\centering
\includegraphics[width=0.29\linewidth]{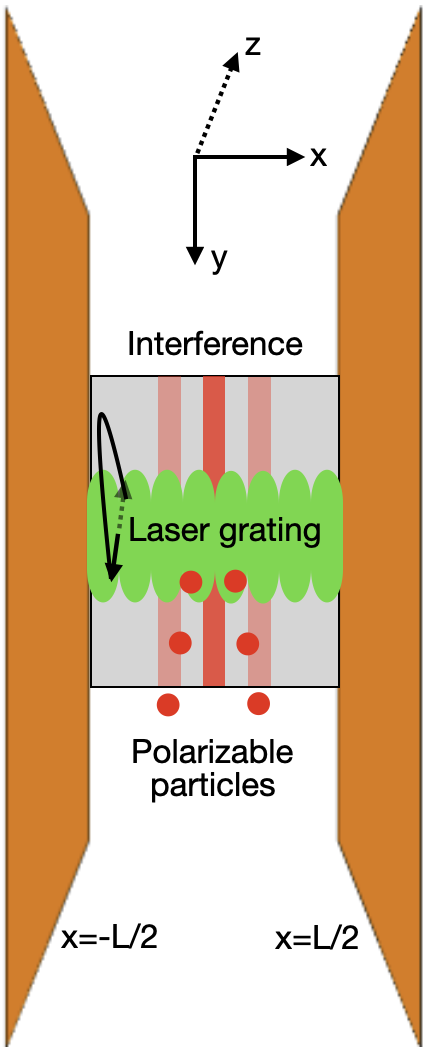}
\caption{{\bf Proposed experimental setup} is shown in the direction of particle (red) propagation toward the laser grating (green). The particle's dipole moment is switched on when it enters the laser grating (having a Gaussian spread along the $z$ axis) at $z\simeq -\sigma_{z}$, and switched off when it exits the grating at $z\simeq\sigma_z$. The same grating  generates a phase difference along the $x$ axis leading to the interference pattern on the screen (gray), partially suppressed by the zero-point modes confined between the conducting plates at $x=\pm L/2$.}\label{Im:ExpSetup}
\end{figure}

 In this Letter we propose a scenario where this can be achieved. The key is to realize that unlike the fixed charge of elementary particles, the dipole moment of neutral polarizable particles can be switched on and off, using an additional laser. We first show that suddenly switching on and off the particle's coupling to the zero-point modes, i.e. the cavity modes in their ground state, leads to an irreversible loss of coherence, and obtain its analytic expression to leading order in the interaction. We then propose an experimental setup (Fig.~\ref{Im:ExpSetup}) that can measure this decoherence effect. (Decoherence for an electron near conducting plates, computed in previous works \cite{Ford1993,Mazzitelli2003,Hsiang2006}, was incorrectly ascribed to VFs \cite{Gundhi2024Casimir}.) Such a measurement would provide a unique probe to measure the spatial and temporal correlations of VFs, providing insights complementing those obtained from the well-known manifestations of the quantum vacuum \cite{BetheLambShift, LambLambShift, Weinberg1989,Fulling,UnruhEffect, Casimir1948, Sabisky1973, Lamoreaux1997, LamoreauxE1998, Lamoreaux1999, Spruch1993,Milton2001,Moore1970,DynCasEffect2011}, and a novel precision test for this fundamental aspect of the decoherence formalism.

\textit{Theory}--A quantitative analysis of decoherence is provided by the reduced density matrix $\hat\rho_r(t)$
\begin{align}\label{AEQ:RedDensityDef}
\hat{\rho}_r(t) = \Sigma_{E}\bra{E}\ket{\Psi_t}\bra{\Psi_t}\ket{E},
\end{align}
where $\ket{\Psi_t}$ denotes the full system-environment (S-E) state and $\ket{E}$  the basis states of the environment. We imagine a situation where the time evolution of the system, encoded by $\psi_t(x)$, is controlled by an external potential. Nevertheless, due to the S-E interaction, the state of the environment $\ket{\mathcal{E}}$ typically becomes correlated to the system position  such that $\ket{\Psi_t} = \int dx\psi_t(x)\ket{\mathcal{E}_x}\ket{x}$, which implies

\begin{align}
\bra{x}\hat\rho_r(t)\ket{x'} = \bra{\mathcal{E}_{x'}}\ket{\mathcal{E}_x}_t\psi_t(x)\psi^{*}_t(x').
\end{align}
The overlap between the environmental states thus motivates the definition of the decoherence kernel $\mathcal{D}$
\begin{align}\label{AEQ:DecKernel}
\mathcal{D}(x,x',t):=\bra{\mathcal{E}_{x'}}\ket{\mathcal{E}_x}_t,
\end{align}
which quantitatively determines the loss in fringe contrast. The S-E correlation, and hence the decoherence kernel, is determined by the interaction Hamiltonian $H_{\text{int}}$. The interaction of a point dipole with an external EM field is described by the Hamiltonian (cf.~p.~271 in \cite{TannoudjiPartOneChapterFour})
\begin{align}\label{AEQ:Hint_p}
H_{\text{int}}=-\mathbf{d}\cdot\mathbf{\Pi},\qquad \mathbf{\Pi}:=-\textbf{P}/\epsilon_0,
\end{align}
where $\textbf{P}$ is the conjugate momentum of the EM field, and the freely evolved Heisenberg operator $\hat{\mathbf{\Pi}}$ is the same as the transverse electric field operator of the free EM field. To leading order in the interaction picture, for a particle located somewhere along the $x$ axis, and for which only the $x$ component of the time-dependent dipole moment $d^x$ is non zero, the state of the environment is given by \begin{align}\label{AEQ:TimeEv}
\ket{\mathcal{E}(x)}_{t} = \exp(\frac{i}{\hbar}\int_{0}^{t} dt'd^{x}(x,t')\hat{\Pi}^{x}(x,t') )\ket{0},
\end{align}
where a factorized initial S-E state ${\hat{\rho}(0) = \hat{\rho}_{\text{S}}(0)\otimes\ket{0}\bra{0}}$, $\ket{0}$ being the vacuum state of the EM field, has been assumed. (Since the dipole moment operator acts trivially on the environmental states, it can be consistently treated as a c-number for computing $\mathcal{D}$.) This assumption is justified since we want to model a scenario in which the dipole moment is switched on (suddenly) at some  time $t>0$, but not before. Given the time evolution~\eqref{AEQ:TimeEv}, the decoherence kernel~\eqref{AEQ:DecKernel} can be computed from the expression for the operator $\hat{\mathbf{\Pi}}$ in the presence of conducting plates \cite{Hsiang2006, Gundhi2024Casimir} (see also the discussion above Eq.~(5.1) in \cite{Barton1991}) 
\begin{align}\label{AEQ:ElecField}
\hat{\mathbf{\Pi}} =& -i\sqrt{\frac{2\hbar}{\epsilon_0L}}\sum_{n=0}^{\infty}f(n)\int \frac{d^2 k_{\parallel}}{2\pi}\sqrt{\frac{\omega_{n}}{2}} e^{i(\k_{\parallel}\cdot \x_{\parallel}-\omega_{n}t)}\nonumber\\
&\times\left[\hat{a}_{1}(\k_{\parallel},n)(\hat{\k}_{\parallel}\times\hat{\x})\sin(n\pi (x+L/2)/L)+ \right.\nonumber\\
&\left.\ \hat{a}_{2}(\k_{\parallel},n)\left\lbrace \frac{i\hat{\k}_{\parallel}n\pi c}{\omega_{n}L}\sin(n\pi (x+L/2)/L)\right.\right.\nonumber\\
&\left.\left.-\frac{\hat{\x} k_{\parallel}c}{\omega_{n}}\cos({n\pi (x+L/2)}/{L})\right\rbrace\right]+\mathrm{H.c.}
\end{align}
Here, $f(0) = 1/\sqrt{2}$ and $f(n)=1\,\forall n>0$, $\hat{\k}_{\parallel}$ is a unit vector along the $y-z$ plane, ${\omega^2_{n}/c^2 := k^2_{\parallel}+n^2\pi^2/L^2}$, and $\hat{a}_1$, $\hat{a}_2$ are the annihilation operators corresponding to the two independent modes of the EM field.

The sudden switching on and off of the dipole moment in our setup can be mathematically modeled by writing $d^x(x,t)=  d(x)s(t)$, where $s(t)$ is the switching function and $d(x)$ the spatial profile of the x-component of the dipole moment. If the dipole moment is switched on and off multiple times, $s(t')$ can be written as
\begin{align}\label{AEQ:Switch}
s(t') =\sum_{m=1}^{N}(-1)^m \theta(t'-t_m),\qquad t_m = m\mathcal{T}\,,
\end{align}
where $\mathcal{T}$ is the time elapsed between switching on and switching off the dipole moment of magnitude $|d(x)|$, and the Heaviside step function models the idealization of a sudden switching on and off of the dipole moment.  Since $\hat{\mathbf{\Pi}}$ is a linear sum of creation and annihilation operators, to leading order in the interaction, $\ket{\mathcal{E}(x)}_{t}$ can be viewed as a coherent state $\ket{\alpha(x,t)}$. This implies that 
\begin{align}\label{AEQ:CoherentState}
&\bra{\alpha(x',t)}\ket{\alpha(x,t)} = \prod_{\k_{\parallel}}\prod_{n}\bra{\alpha_{\k_{\parallel} n}(x',t)}\ket{\alpha_{\k_{\parallel} n}(x,t)}\,,
\end{align}
can be evaluated using the identity $\bra{\alpha_1}\ket{\alpha_2} = e^{-(|\alpha_1|^2+|\alpha_2|^2-2\alpha_1^*\alpha_2)/2}$, as detailed in the End Matter. To compute $\mathcal{D}$, we consider particle superpositions near the center $x=0$. Such a scenario is also more feasible, as it would be free of unwanted image effects that become relevant close to the conducting plates. See, for example, Appendix C of the End Matter, where we show that decoherence due to image currents \cite{Anglin1996} would be insignificant near $x=0$. 

\begin{figure}[!ht]
\centering
\includegraphics[width=1.0\linewidth]{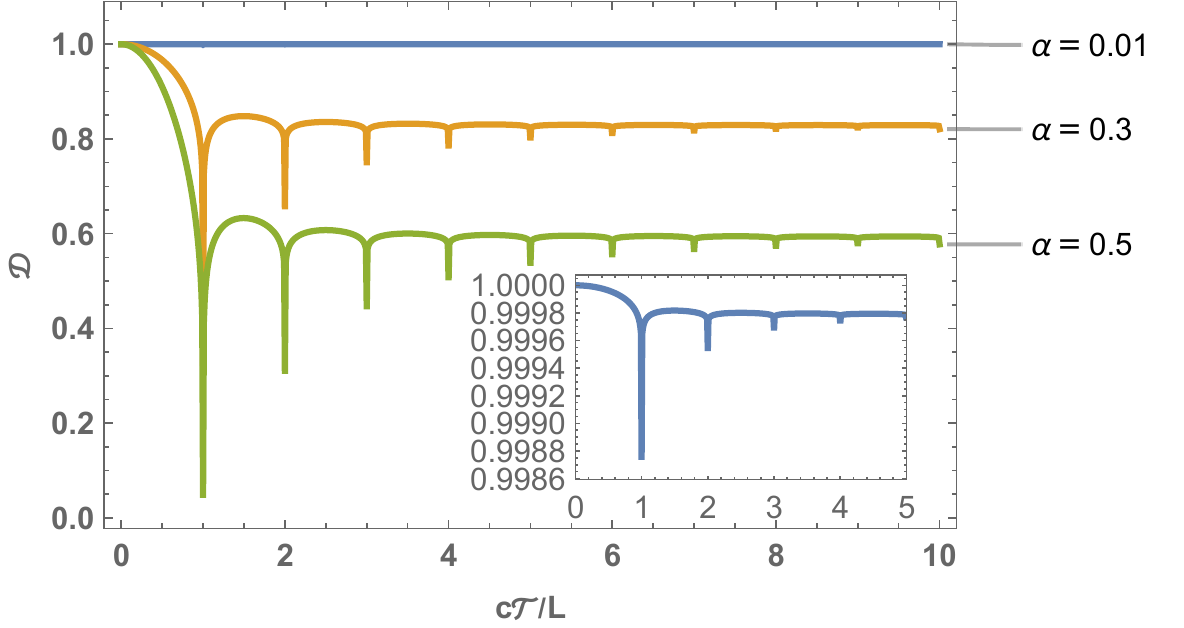}
\caption{{\bf Decoherence} is shown as a function of $\mathcal{T}$ (the time elapsed between the sudden switching on and off of the interaction) and as a function of $\alpha$, obtained by setting $\kappa=10^{8}$ in Eq.~\eqref{AEQ:DecCutoff}. As per  Eq.~\eqref{AEQ:DecKernel}, $\mathcal{D}=1$ implies no decoherence and $\mathcal{D}=0$ signifies maximum decoherence.}\label{Im:DecPlot}
\end{figure}

 In our setup, the dipole moment of the particles is switched on while they interact with the additional laser confined between $-\sigma_z\lesssim z\lesssim \sigma_z$, as depicted in Fig.~\ref{Im:ExpSetup}. The spatial variation of $d(x)$ is therefore controlled by the spatial profile of the laser, and not the zero-point modes. This is because, typically, the dipole moment of neutral polarizable molecules is insignificant in empty space. It is enhanced by the molecule-laser interaction in our proposal. Further, we assume that the superposition is prepared over length scales that are much smaller than $L$, but comparable to the wavelength of the laser. In such a scenario, the state of the environment can be computed within the so-called dipole approximation, where the spatial variation of the zero-point modes can be ignored, such that
\begin{align}\label{AEQ:Dipole}
\ket{\mathcal{E}(x)}_{t} \approx \exp(\frac{i d(x)}{\hbar}\int_{0}^{t} dt's(t')\hat{\Pi}^{x}(0,t') )\ket{0}.
\end{align}
Within the dipole approximation, as detailed in the End Matter, for $t>t_N$, $\mathcal{D}(x,x',t)$ is obtained to be
\begin{align}\label{AEQ:OverlapAdiabatic}
\mathcal{D}=&\bra{\mathcal{E}(x')}\ket{\mathcal{E}(x)}_{t>t_N} =\exp\left\lbrace\frac{-(d(x')-d(x))^2}{4\pi^2\hbar \epsilon_0L}\sum_{n=-\infty}^{\infty}\right.\nonumber\\ 
&\left.\times\int d\k_{\parallel}\frac{(k_{\parallel }c)^2\cos^2(n\pi/2)}{ 4\omega^3_{n}}\frac{\sin^2(N \omega_n\mathcal{T}/2)}{\cos^2(\omega_n\mathcal{T}/2)}\right\rbrace\,.
\end{align}
Next, we set $N=2$ since in our proposed setup the dipole moment is switched on and off only once and not multiple times. Then, by writing $\int d\k_{\parallel}$ as $\int d\k \delta_n(k_x)$, and using the identity 
\begin{align}\label{AEQ:DiracComb}
\sum_{n\text{ even}} \delta_{n} = \frac{L}{2\pi}\sum_{m=-\infty}^{\infty} e^{imk_x L},\qquad \delta_n:=\delta\left(k_x-\frac{n\pi}{L}\right),
\end{align}
the integral in Eq.~\eqref{AEQ:OverlapAdiabatic} can be evaluated as shown in Appendix A of the End Matter. See also Appendix B in the End Matter, where it is shown for a special case that the functional form of $\mathcal{D}$ does not necessarily change even when the dipole approximation is not applied. In order to be consistent with the point particle treatment of the molecules, the integral is computed after imposing a cutoff in the radial part of the integral $\int d\k \to\int_{0}^{k_{\mathrm{max}}} dk  k^2\int d\theta  d\phi\sin\theta $. Evaluating the integral~\eqref{AEQ:OverlapAdiabatic} with the cutoff gives \footnote{In the absence of a cutoff, the decoherence kernel takes the  compact form $\mathcal{D}=\exp{\frac{-(d(x')-d(x))^2}{2\pi^2\hbar\epsilon_0 L^3}\sum_{m=1}^{\infty}\frac{\mathcal{T}}{m^3}\ln(\left|\frac{mL+c\mathcal{T}}{mL-c\mathcal{T}}\right|)}$. Nevertheless, physically, $\mathcal{D}$ must be obtained by setting a finite cutoff as explained in the main text.}
\begin{widetext}
\begin{align}\label{AEQ:DecCutoff}
\mathcal{D}=\exp \left\lbrace\-\sum _{m=1}^{\infty} \frac{-2\alpha^2(x,x')}{m^3\pi\kappa}\left(\kappa  \tau  \left[\ln\left(\left|\frac{m+\tau}{m-\tau}\right|\right)+\text{Ci}(\kappa |m-\tau |)-\text{Ci}\left(\kappa (m+\tau )\right)\right]-4 \sin^2 \left(\frac{\kappa  \tau}{2} \right) \sin (m\kappa)\right)\right\rbrace.
\end{align}
\end{widetext}
Here, $\alpha^2(x,x'):=(d(x')-d(x))^2/(4\pi\epsilon_0\hbar c L^2)$ can be viewed as the dipole moment version of the fine-structure ``constant", $\kappa:= k_{\mathrm{max}}L$ and $\tau:=c\mathcal{T}/L$ are dimensionless parameters, and Ci is the cosine integral function. The dependence of decoherence on $\alpha$ and the switching off time $\mathcal{T}$ is shown in Fig.~\ref{Im:DecPlot}.

As detailed in the End Matter, in Eq.~\eqref{AEQ:DecCutoff}, we have left out the $m=0$ term, since it captures the effect of VFs without boundaries \cite{Gundhi2024Casimir} and corresponds to false decoherence \cite{Gundhi:2023vjs} in the scenario that we are proposing in this Letter. 

As we will explain shortly, the optimal separation between the plates is expected to be $L\simeq 10^{-3}\text{m}-10^{-2}\text{m}$ for decoherence to be observable. For this value, $\kappa=10^{8}$ set in Fig.~\ref{Im:DecPlot} implies that the contribution of mode wavelengths shorter than $\lambda_{\mathrm{min}}=1/k_{\mathrm{max}}\simeq 10^{-10}\text{m}$ to $\mathcal{D}$  is excluded. This cutoff is physically motivated from the molecule sizes  $a\simeq 10^{-10}-10^{-9}$m, for which we propose the experiment to be performed.  Doing so we only retain zero-point modes with wavelengths larger than the size of the molecules, in order to be consistent with the point-particle treatment of molecules. Moreover, the numerical value of the cutoff only affects the value of $\mathcal{D}$ in the close proximity of $\mathcal{T}=mL/c$, but not at late times. It is not clear if it would be feasible to experimentally detect the sharp falls at $\mathcal{T}=mL/c$, since it would require high control over the time spent by the particles in the grating. Nevertheless, the late time behavior can still be tested experimentally, which is not sensitive to the precise numerical value of the cutoff, as long as it is sufficiently large. 

We emphasize that the $m=0$ term and/or the short wavelength modes might only add to  decoherence. By removing their contribution from the decoherence kernel~\eqref{AEQ:DecCutoff}, we obtain a more conservative (and realistic) estimate for decoherence one might detect experimentally.

\textit{Experimental proposal}--A promising scenario for the experimental realization of this proposal is to adapt the settings of molecule interferometry as pioneered by the Arndt group in Vienna~\cite{hornberger2012colloquium, juffmann2013experimental}, and especially the configurations with phase gratings made of laser light as the generator of particle's spatial quantum superpositions, namely the Kapitza-Dirac-Talbot-Lau Interferometer (KDTLI)~\cite{gerlich2007kapitza} and Optical Time-domain Ionizing Matter-wave (OTIMA) configurations~\cite{haslinger2013universal}.  

The experimental setup is sketched in Fig.~\ref{Im:ExpSetup}. The laser grating  at Ultra-violet wavelength, depicted in green, is arranged orthogonal to the propagation direction of the particles. Typically, the intensity profile of the laser grating is given by 
\begin{align}\label{AEQ:Laser}
I(x,y,z)=\frac{8P}{\pi \sigma_z\sigma_y}\exp{-\frac{2y^2}{\sigma^2_y}-\frac{2z^2}{\sigma^2_z}}\sin^2(\pi x/l),
\end{align}
where $\sigma_y$ and $\sigma_z$ denote the \textit{spread} of the laser along the $y$ and the $z$ axis respectively, $P$ is the power, and $l$ the \textit{grating} separation generated by the laser profile along the $x$-axis.

The laser grating serves two purposes. First, it generates a phase difference along the $x$-axis as per the Kapitza-Dirac effect~\cite{kapitza1933reflection,Hornberger2009}, leading to the interference pattern depicted in red in Fig.~\ref{Im:ExpSetup}. Intuitively, this is because particles at different coordinates face a different intensity $I(x,y,z)$, and thus a different effective potential $V(x,y,z)$  \cite{Hornberger2009}. This leads to the position dependent phase \cite{Hornberger2009}
\begin{align}\label{AEQ:phase}
\varphi(x) = \varphi_0\sin^2(\pi x/l)\,,\qquad \varphi_0:=8\sqrt{2\pi}\frac{\alpha_{p}}{\hbar c}\frac{P}{\sigma_y v_z}.
\end{align}
Second, the grating switches on and off the dipole moment of the particles as they enter ($z\simeq-\sigma_{z}$) and leave ($z\simeq\sigma_{z}$) the grating respectively. This, consequently, switches on and off their interaction with the vacuum fluctuations (Eq.~\eqref{AEQ:Hint_p}). This mechanism is essential to detect decoherence due to VFs, which is otherwise difficult to achieve with elementary particles having a fixed charge. 

As can be seen from Eq.~\eqref{AEQ:DecCutoff} and Fig.~\ref{Im:DecPlot}, decoherence depends on the polarizability of the particles, separation between the plates $L$, and also on the time spent by the particles in the grating $\mathcal{T}$. It further depends, as we describe shortly, on the laser power $P$ and $\sigma_{z}$, as they determine the magnitude of the dipole moment generated by the molecule-laser interaction. 

The question then arises of whether there exist molecules that have a polarizability high enough for which the interference experiment can be performed and this decoherence effect detected. Typically, a $5\%-10\%$ relative change in visibility $\mathcal{V}$ of the interference pattern is experimentally detectable \cite{ArndtCollDec2003}. While a detailed computation of visibility under this decoherence effect, in an actual Talbot-Lau setup, deserves an independent analysis and is outside the scope of the present Letter, we expect the relative change in visibility to scale with $1-\mathcal{D}$ (cf.~Eqs.~(55) and ~(71) in Ref.~\cite{Hornberger2004}). Thus, for certain molecules for which the dipole moment generated by the laser grating brings them close to the green curve in Fig.~\ref{Im:DecPlot} ($\alpha=0.5$), compared to those with a much weaker polarizability, or for laser parameters for which decoherence is expected to be negligible, a relative change in visibility of the order of 40\% is expected to be achieved in an ideal scenario, which is within the reach of an experimental detection. 

We emphasize that this estimate for visibility must be confirmed by a future work, dedicated to quantifying this novel decoherence effect within the Talbot-Lau framework \cite{Hornberger2004}. Below we estimate the dipole moment that can be generated for C$_{60}$ molecules and sodium clusters on the mass range of $10^{6}$ atomic mass units (amu), and compare them to the value required for an experimentally detectable loss in the visibility of the pattern.

\textit{Estimates and experimental feasibility: The plate separation}--As emphasized in the introduction, in order to observe decoherence due to VFs, the dipole moment needs to be switched on and off suddenly. In our proposed setup, suddenly means much faster than $L/c$ -- the characteristic timescale of the environment.  The time it takes for the dipole moment to switch on is the time it takes the particle to enter the grating, i.e., to traverse its own diameter $a\simeq 10^{-9}-10^{-10}$m. Although the particles move much slower than light ($v_z/c\simeq 10^{-6}$ \cite{hackermuller2004decoherence}), the sudden switching on and off can still be achieved if $L$ is much bigger than the particle size. We take $L\simeq 10^{-3}-10^{-2}$m for our proposal, which is sufficient to implement a sharp interaction switch since 
\begin{align}\label{AEQ:SwitchOnOff}
    a/L\simeq 10^{-8}-10^{-6}\lesssim v_z/c\simeq 10^{-6}.
\end{align}
The relation above restricts the desirable value of $L$ to a very specific range, as larger values of $L$ reduce the decoherence effect while for shorter values, $a/L$ will not be sufficiently small for the interaction to be switched on and off nonadiabatically.

\textit{The dipole moment}--The orange and green curves in Fig.~\ref{Im:DecPlot} show that decoherence becomes detectable for $\alpha$ greater than some critical value $\alpha_{\mathrm{crit}}\simeq 0.1-0.5$. This implies that the laser-molecule interaction must be able to generate a dipole moment of
\begin{align}\label{AEQ:DipoleWanted}
|\mathbf{d}| \gtrsim \alpha_{\mathrm{crit}} L (4\pi\epsilon_0\hbar c )^{1/2}\simeq10^{-21}-10^{-22}\,\text{C}\,\text{m},
\end{align}
for $L$ ranging between $L\simeq 10^{-2}-10^{-3}$m. Molecules are typically characterized by their polarizability $\alpha_p$ which is related to the dipole moment as $|\mathbf{d}| = \alpha_p|\mathbf{E}|$, $|\mathbf{E}|$ being the electric field amplitude (in our case, of the laser grating). The dipole moment $|\mathbf{d}|$ generated for a given molecule with a given polarizability $\alpha_p$, can be estimated by obtaining  $|\mathbf{E}|$ from the laser power $P$, using the standard relation $|\mathbf{E}| = \sqrt{2I/(c\epsilon_0)}\simeq\sqrt{16 P/(\pi\sigma_z\sigma_y\epsilon_0 c)}$. In order to generate the maximum dipole moment, we can reduce $\sigma_z$ all the way down to the diffraction limit \cite{Novotny_Hecht_2012} $\sigma_z\simeq l\simeq 10^{-7}$ m, where $l$ is the wavelength of the laser grating in Eq.~\eqref{AEQ:Laser}. Doing so would have no side effects, for instance, for the phase $\varphi$ generated by the grating along the $x$ axis (cf.~Eq.~\eqref{AEQ:phase}), as it does not depend on $\sigma_z$ \cite{Hornberger2009}. Thus, taking $\sigma_z = 10^{-7}$ m and $\sigma_y = 10^{-3}\,\text{m}$, $P=10\,\text{W}$ as in \cite{Hornberger2009}, we get $|\mathbf{E}| \simeq 10^{6}-10^{7}\,\text{V/m}$. 

We now consider two different molecules. These are the C$_{60}$ molecules, and sodium clusters on the mass range of $10^6$ amu. Their respective polarizabilities are $\alpha^{60}_p \simeq 10^{-32} \text{C}\,\text{m}^2\,\text{V}^{-1}$ \cite{Arndt2007Pol,PolarizationBook2011} and $\alpha^{\text{Na}}_p\simeq 10^{-29} \text{C}\,\text{m}^2\,\text{V}^{-1}$  \cite{Kresin2001,PolarizationBook2011}. Using the electric field amplitude of the laser calculated above, we get that the dipole moment $|\mathbf{d}_{60}|$ that can be generated for C$_{60}$ molecules by the molecule-laser interaction is
\begin{align}\label{AEQ:dipoleC60}
|\mathbf{d}_{60}|\simeq 10^{-25} \text{C}\,\text{m}\ll 10^{-22}  \text{C}\,\text{m}.
\end{align}
Instead, for sodium clusters on the mass range of $10^6$ amu, the dipole moment $|\mathbf{d}_{\text{Na}}|$ is given by
\begin{align}\label{AEQ:dipoleNa}
|\mathbf{d}_{\text{Na}}|\simeq 10^{-22}\text{C}\,\text{m}.
\end{align}
Therefore, we have shown that while decoherence due to vacuum fluctuations would be difficult to detect for the commonly used C$_{60}$ molecules (Eq.~\eqref{AEQ:dipoleC60}), since the dipole moment generated for C$_{60}$ is short from the desired value~\eqref{AEQ:DipoleWanted} by 3 to 4 orders of magnitude, it can be detected by performing interference experiments with alkali earth clusters (Eq.~\eqref{AEQ:dipoleNa}) instead. The possibility of detecting this effect with the latter is particularly interesting, since matter-wave interferometry for sodium clusters is already being developed at Vienna~\cite{MarkusPrivate}.

\textit{Conclusions}.-- In this Letter our aim is to show that decoherence due to vacuum fluctuations, requiring the interaction between the system and the zero-point modes of the vacuum to be suddenly switched on and off, can be measured with the help of modern interferometric experiments when performed with alkali earth clusters. Prediction of the decoherence effect in Eq.~\eqref{AEQ:DecCutoff} can be tested by varying the laser power, the plate separation, and the spatial profile of the laser, all of which lead to a continuous change in decoherence. It is not unlikely for  decoherence due to vacuum fluctuations to be measured in the near future, given that Talbot-Lau matter-wave interferometry with sodium clusters in the mass range of $10^6$ amu is being developed at Vienna, and preliminary results have been achieved very recently~\cite{MarkusPrivate}. 

Such a measurement would probe the foundations of quantum field theory and the decoherence formalism, and might also be relevant for the recent proposals to test the quantum nature of gravity \cite{MarlettoVedral2017,Milburn2017}. This is because an experimental realization of our proposal would bring to light a fundamentally unavoidable source of decoherence, which has not been discussed previously, and that might become dominant for massive/highly polarizable particles in standard matter-wave interferometry.

 \textit{Acknowledgments}--This work is financially supported by the University of Trieste, INFN and EU Horizon Europe EIC Pathfinder project QuCoM (GA No. 101046973 and No. 10032223). We would further like to acknowledge support the UKRI EPSRC (EP/W007444/1, EP/V000624/1, EP/V035975/1 and EP/X009491/1), the Leverhulme Trust (RPG-2022-57) and the QuantERA II Programme (project LEMAQUME), which has received funding from the European Union’s Horizon 2020 research and innovation programme under Grant Agreement No 101017733.

 \textit{Data availability}--No data were created or analyzed in this study.
\bibliography{VacuumDecoherenceBib}{}

\begin{thebibliography}{52}%
\makeatletter
\providecommand \@ifxundefined [1]{%
 \@ifx{#1\undefined}
}%
\providecommand \@ifnum [1]{%
 \ifnum #1\expandafter \@firstoftwo
 \else \expandafter \@secondoftwo
 \fi
}%
\providecommand \@ifx [1]{%
 \ifx #1\expandafter \@firstoftwo
 \else \expandafter \@secondoftwo
 \fi
}%
\providecommand \natexlab [1]{#1}%
\providecommand \enquote  [1]{``#1''}%
\providecommand \bibnamefont  [1]{#1}%
\providecommand \bibfnamefont [1]{#1}%
\providecommand \citenamefont [1]{#1}%
\providecommand \href@noop [0]{\@secondoftwo}%
\providecommand \href [0]{\begingroup \@sanitize@url \@href}%
\providecommand \@href[1]{\@@startlink{#1}\@@href}%
\providecommand \@@href[1]{\endgroup#1\@@endlink}%
\providecommand \@sanitize@url [0]{\catcode `\\12\catcode `\$12\catcode
  `\&12\catcode `\#12\catcode `\^12\catcode `\_12\catcode `\%12\relax}%
\providecommand \@@startlink[1]{}%
\providecommand \@@endlink[0]{}%
\providecommand \url  [0]{\begingroup\@sanitize@url \@url }%
\providecommand \@url [1]{\endgroup\@href {#1}{\urlprefix }}%
\providecommand \urlprefix  [0]{URL }%
\providecommand \Eprint [0]{\href }%
\providecommand \doibase [0]{https://doi.org/}%
\providecommand \selectlanguage [0]{\@gobble}%
\providecommand \bibinfo  [0]{\@secondoftwo}%
\providecommand \bibfield  [0]{\@secondoftwo}%
\providecommand \translation [1]{[#1]}%
\providecommand \BibitemOpen [0]{}%
\providecommand \bibitemStop [0]{}%
\providecommand \bibitemNoStop [0]{.\EOS\space}%
\providecommand \EOS [0]{\spacefactor3000\relax}%
\providecommand \BibitemShut  [1]{\csname bibitem#1\endcsname}%
\let\auto@bib@innerbib\@empty
%</preamble>
\bibitem [{\citenamefont {Barone}\ and\ \citenamefont
  {Caldeira}(1991)}]{Caldeira1991}%
  \BibitemOpen
  \bibfield  {author} {\bibinfo {author} {\bibfnamefont {P.~M. V.~B.}\
  \bibnamefont {Barone}}\ and\ \bibinfo {author} {\bibfnamefont {A.~O.}\
  \bibnamefont {Caldeira}},\ }\bibfield  {title} {\bibinfo {title} {Quantum
  mechanics of radiation damping},\ }\href
  {https://doi.org/10.1103/PhysRevA.43.57} {\bibfield  {journal} {\bibinfo
  {journal} {Phys. Rev. A}\ }\textbf {\bibinfo {volume} {43}},\ \bibinfo
  {pages} {57} (\bibinfo {year} {1991})}\BibitemShut {NoStop}%
\bibitem [{\citenamefont {Kiefer}(1992)}]{Kiefer1992}%
  \BibitemOpen
  \bibfield  {author} {\bibinfo {author} {\bibfnamefont {C.}~\bibnamefont
  {Kiefer}},\ }\bibfield  {title} {\bibinfo {title} {Decoherence in quantum
  electrodynamics and quantum gravity},\ }\href
  {https://doi.org/10.1103/PhysRevD.46.1658} {\bibfield  {journal} {\bibinfo
  {journal} {Phys. Rev. D}\ }\textbf {\bibinfo {volume} {46}},\ \bibinfo
  {pages} {1658} (\bibinfo {year} {1992})}\BibitemShut {NoStop}%
\bibitem [{\citenamefont {Ford}(1993)}]{Ford1993}%
  \BibitemOpen
  \bibfield  {author} {\bibinfo {author} {\bibfnamefont {L.~H.}\ \bibnamefont
  {Ford}},\ }\bibfield  {title} {\bibinfo {title} {Electromagnetic vacuum
  fluctuations and electron coherence},\ }\href
  {https://doi.org/10.1103/PhysRevD.47.5571} {\bibfield  {journal} {\bibinfo
  {journal} {Phys. Rev. D}\ }\textbf {\bibinfo {volume} {47}},\ \bibinfo
  {pages} {5571} (\bibinfo {year} {1993})}\BibitemShut {NoStop}%
\bibitem [{\citenamefont {Santos}(1994)}]{Santos1994}%
  \BibitemOpen
  \bibfield  {author} {\bibinfo {author} {\bibfnamefont {E.}~\bibnamefont
  {Santos}},\ }\bibfield  {title} {\bibinfo {title} {Objectification of
  classical properties induced by quantum vacuum fluctuations},\ }\href
  {https://doi.org/https://doi.org/10.1016/0375-9601(94)90438-3} {\bibfield
  {journal} {\bibinfo  {journal} {Physics Letters A}\ }\textbf {\bibinfo
  {volume} {188}},\ \bibinfo {pages} {198} (\bibinfo {year}
  {1994})}\BibitemShut {NoStop}%
\bibitem [{\citenamefont {Di\'{o}si}(1995)}]{Diosi1995}%
  \BibitemOpen
  \bibfield  {author} {\bibinfo {author} {\bibfnamefont {L.}~\bibnamefont
  {Di\'{o}si}},\ }\bibfield  {title} {\bibinfo {title} {Comments on
  “objectification of classical properties induced by quantum vacuum
  fluctuations”},\ }\href
  {https://doi.org/https://doi.org/10.1016/0375-9601(94)00846-H} {\bibfield
  {journal} {\bibinfo  {journal} {Physics Letters A}\ }\textbf {\bibinfo
  {volume} {197}},\ \bibinfo {pages} {183} (\bibinfo {year}
  {1995})}\BibitemShut {NoStop}%
\bibitem [{\citenamefont {Elze}(1995)}]{ELZE1995}%
  \BibitemOpen
  \bibfield  {author} {\bibinfo {author} {\bibfnamefont {H.-T.}\ \bibnamefont
  {Elze}},\ }\bibfield  {title} {\bibinfo {title} {Vacuum-induced quantum
  decoherence and the entropy puzzle},\ }\href
  {https://doi.org/https://doi.org/10.1016/0920-5632(95)00066-I} {\bibfield
  {journal} {\bibinfo  {journal} {Nuclear Physics B - Proceedings Supplements}\
  }\textbf {\bibinfo {volume} {39}},\ \bibinfo {pages} {169} (\bibinfo {year}
  {1995})}\BibitemShut {NoStop}%
\bibitem [{\citenamefont {Breuer}\ and\ \citenamefont
  {Petruccione}(2000)}]{BandP}%
  \BibitemOpen
  \bibfield  {author} {\bibinfo {author} {\bibfnamefont {H.-P.}\ \bibnamefont
  {Breuer}}\ and\ \bibinfo {author} {\bibfnamefont {F.}~\bibnamefont
  {Petruccione}},\ }\bibfield  {title} {\bibinfo {title} {Radiation damping and
  decoherence in quantum electrodynamics},\ }in\ \href
  {https://link.springer.com/chapter/10.1007/3-540-45369-5_3} {\emph {\bibinfo
  {booktitle} {Relativistic Quantum Measurement and Decoherence}}},\ \bibinfo
  {editor} {edited by\ \bibinfo {editor} {\bibfnamefont {H.-P.}\ \bibnamefont
  {Breuer}}\ and\ \bibinfo {editor} {\bibfnamefont {F.}~\bibnamefont
  {Petruccione}}}\ (\bibinfo  {publisher} {Springer},\ \bibinfo {address}
  {Berlin, Heidelberg},\ \bibinfo {year} {2000})\ pp.\ \bibinfo {pages}
  {31--65}\BibitemShut {NoStop}%
\bibitem [{\citenamefont {Unruh}(2000)}]{Unruh_Coherence}%
  \BibitemOpen
  \bibfield  {author} {\bibinfo {author} {\bibfnamefont {W.~G.}\ \bibnamefont
  {Unruh}},\ }\bibfield  {title} {\bibinfo {title} {False loss of coherence},\
  }in\ \href {https://link.springer.com/chapter/10.1007/3-540-45369-5_7} {\emph
  {\bibinfo {booktitle} {Relativistic Quantum Measurement and Decoherence}}},\
  \bibinfo {editor} {edited by\ \bibinfo {editor} {\bibfnamefont {H.-P.}\
  \bibnamefont {Breuer}}\ and\ \bibinfo {editor} {\bibfnamefont
  {F.}~\bibnamefont {Petruccione}}}\ (\bibinfo  {publisher} {Springer},\
  \bibinfo {address} {Berlin, Heidelberg},\ \bibinfo {year} {2000})\ pp.\
  \bibinfo {pages} {125--140}\BibitemShut {NoStop}%
\bibitem [{\citenamefont {Mazzitelli}\ \emph {et~al.}(2003)\citenamefont
  {Mazzitelli}, \citenamefont {Paz},\ and\ \citenamefont
  {Villanueva}}]{Mazzitelli2003}%
  \BibitemOpen
  \bibfield  {author} {\bibinfo {author} {\bibfnamefont {F.~D.}\ \bibnamefont
  {Mazzitelli}}, \bibinfo {author} {\bibfnamefont {J.~P.}\ \bibnamefont
  {Paz}},\ and\ \bibinfo {author} {\bibfnamefont {A.}~\bibnamefont
  {Villanueva}},\ }\bibfield  {title} {\bibinfo {title} {Decoherence and
  recoherence from vacuum fluctuations near a conducting plate},\ }\href
  {https://doi.org/10.1103/PhysRevA.68.062106} {\bibfield  {journal} {\bibinfo
  {journal} {Phys. Rev. A}\ }\textbf {\bibinfo {volume} {68}},\ \bibinfo
  {pages} {062106} (\bibinfo {year} {2003})}\BibitemShut {NoStop}%
\bibitem [{\citenamefont {Hsiang}\ and\ \citenamefont
  {Lee}(2006)}]{Hsiang2006}%
  \BibitemOpen
  \bibfield  {author} {\bibinfo {author} {\bibfnamefont {J.-T.}\ \bibnamefont
  {Hsiang}}\ and\ \bibinfo {author} {\bibfnamefont {D.-S.}\ \bibnamefont
  {Lee}},\ }\bibfield  {title} {\bibinfo {title} {Influence on electron
  coherence from quantum electromagnetic fields in the presence of conducting
  plates},\ }\href {https://doi.org/10.1103/PhysRevD.73.065022} {\bibfield
  {journal} {\bibinfo  {journal} {Phys. Rev. D}\ }\textbf {\bibinfo {volume}
  {73}},\ \bibinfo {pages} {065022} (\bibinfo {year} {2006})}\BibitemShut
  {NoStop}%
\bibitem [{\citenamefont {Baym}\ and\ \citenamefont
  {Ozawa}(2009)}]{Baym_Ozawa}%
  \BibitemOpen
  \bibfield  {author} {\bibinfo {author} {\bibfnamefont {G.}~\bibnamefont
  {Baym}}\ and\ \bibinfo {author} {\bibfnamefont {T.}~\bibnamefont {Ozawa}},\
  }\bibfield  {title} {\bibinfo {title} {Two-slit diffraction with highly
  charged particles: Niels bohr's consistency argument that the electromagnetic
  field must be quantized},\ }\href {https://doi.org/10.1073/pnas.0813239106}
  {\bibfield  {journal} {\bibinfo  {journal} {Proc. Natl. Acad. Sci. U.S.A.}\
  }\textbf {\bibinfo {volume} {106}},\ \bibinfo {pages} {3035} (\bibinfo {year}
  {2009})}\BibitemShut {NoStop}%
\bibitem [{\citenamefont {Gundhi}\ and\ \citenamefont
  {Bassi}(2023)}]{Gundhi:2023vjs}%
  \BibitemOpen
  \bibfield  {author} {\bibinfo {author} {\bibfnamefont {A.}~\bibnamefont
  {Gundhi}}\ and\ \bibinfo {author} {\bibfnamefont {A.}~\bibnamefont {Bassi}},\
  }\bibfield  {title} {\bibinfo {title} {{Motion of an electron through vacuum
  fluctuations}},\ }\href {https://doi.org/10.1103/PhysRevA.107.062801}
  {\bibfield  {journal} {\bibinfo  {journal} {Phys. Rev. A}\ }\textbf {\bibinfo
  {volume} {107}},\ \bibinfo {pages} {062801} (\bibinfo {year}
  {2023})}\BibitemShut {NoStop}%
\bibitem [{\citenamefont {Gundhi}(2024)}]{Gundhi2024Casimir}%
  \BibitemOpen
  \bibfield  {author} {\bibinfo {author} {\bibfnamefont {A.}~\bibnamefont
  {Gundhi}},\ }\bibfield  {title} {\bibinfo {title} {Decoherence due to the
  casimir effect?},\ }\href {https://doi.org/10.1103/PhysRevD.110.116001}
  {\bibfield  {journal} {\bibinfo  {journal} {Phys. Rev. D}\ }\textbf {\bibinfo
  {volume} {110}},\ \bibinfo {pages} {116001} (\bibinfo {year}
  {2024})}\BibitemShut {NoStop}%
\bibitem [{\citenamefont {Kiefer}(2003)}]{Kiefer2003}%
  \BibitemOpen
  \bibfield  {author} {\bibinfo {author} {\bibfnamefont {C.}~\bibnamefont
  {Kiefer}},\ }\bibinfo {title} {Decoherence in quantum field theory and
  quantum gravity},\ in\ \href {https://doi.org/10.1007/978-3-662-05328-7_4}
  {\emph {\bibinfo {booktitle} {Decoherence and the Appearance of a Classical
  World in Quantum Theory}}}\ (\bibinfo  {publisher} {Springer},\ \bibinfo
  {address} {Berlin, Heidelberg},\ \bibinfo {year} {2003})\ pp.\ \bibinfo
  {pages} {181--225}\BibitemShut {NoStop}%
\bibitem [{\citenamefont {Bethe}(1947)}]{BetheLambShift}%
  \BibitemOpen
  \bibfield  {author} {\bibinfo {author} {\bibfnamefont {H.~A.}\ \bibnamefont
  {Bethe}},\ }\bibfield  {title} {\bibinfo {title} {The electromagnetic shift
  of energy levels},\ }\href {https://doi.org/10.1103/PhysRev.72.339}
  {\bibfield  {journal} {\bibinfo  {journal} {Phys. Rev.}\ }\textbf {\bibinfo
  {volume} {72}},\ \bibinfo {pages} {339} (\bibinfo {year} {1947})}\BibitemShut
  {NoStop}%
\bibitem [{\citenamefont {Lamb}\ and\ \citenamefont
  {Retherford}(1947)}]{LambLambShift}%
  \BibitemOpen
  \bibfield  {author} {\bibinfo {author} {\bibfnamefont {W.~E.}\ \bibnamefont
  {Lamb}}\ and\ \bibinfo {author} {\bibfnamefont {R.~C.}\ \bibnamefont
  {Retherford}},\ }\bibfield  {title} {\bibinfo {title} {Fine structure of the
  hydrogen atom by a microwave method},\ }\href
  {https://doi.org/10.1103/PhysRev.72.241} {\bibfield  {journal} {\bibinfo
  {journal} {Phys. Rev.}\ }\textbf {\bibinfo {volume} {72}},\ \bibinfo {pages}
  {241} (\bibinfo {year} {1947})}\BibitemShut {NoStop}%
\bibitem [{\citenamefont {Weinberg}(1989)}]{Weinberg1989}%
  \BibitemOpen
  \bibfield  {author} {\bibinfo {author} {\bibfnamefont {S.}~\bibnamefont
  {Weinberg}},\ }\bibfield  {title} {\bibinfo {title} {The cosmological
  constant problem},\ }\href {https://doi.org/10.1103/RevModPhys.61.1}
  {\bibfield  {journal} {\bibinfo  {journal} {Rev. Mod. Phys.}\ }\textbf
  {\bibinfo {volume} {61}},\ \bibinfo {pages} {1} (\bibinfo {year}
  {1989})}\BibitemShut {NoStop}%
\bibitem [{\citenamefont {Fulling}(1973)}]{Fulling}%
  \BibitemOpen
  \bibfield  {author} {\bibinfo {author} {\bibfnamefont {S.~A.}\ \bibnamefont
  {Fulling}},\ }\bibfield  {title} {\bibinfo {title} {Nonuniqueness of
  canonical field quantization in riemannian space-time},\ }\href
  {https://doi.org/10.1103/PhysRevD.7.2850} {\bibfield  {journal} {\bibinfo
  {journal} {Phys. Rev. D}\ }\textbf {\bibinfo {volume} {7}},\ \bibinfo {pages}
  {2850} (\bibinfo {year} {1973})}\BibitemShut {NoStop}%
\bibitem [{\citenamefont {Unruh}(1976)}]{UnruhEffect}%
  \BibitemOpen
  \bibfield  {author} {\bibinfo {author} {\bibfnamefont {W.~G.}\ \bibnamefont
  {Unruh}},\ }\bibfield  {title} {\bibinfo {title} {Notes on black-hole
  evaporation},\ }\href {https://doi.org/10.1103/PhysRevD.14.870} {\bibfield
  {journal} {\bibinfo  {journal} {Phys. Rev. D}\ }\textbf {\bibinfo {volume}
  {14}},\ \bibinfo {pages} {870} (\bibinfo {year} {1976})}\BibitemShut
  {NoStop}%
\bibitem [{\citenamefont {Casimir}(1948)}]{Casimir1948}%
  \BibitemOpen
  \bibfield  {author} {\bibinfo {author} {\bibfnamefont {H.~B.~G.}\
  \bibnamefont {Casimir}},\ }\bibfield  {title} {\bibinfo {title} {{On the
  Attraction Between Two Perfectly Conducting Plates}},\ }\href@noop {}
  {\bibfield  {journal} {\bibinfo  {journal} {Indag. Math.}\ }\textbf {\bibinfo
  {volume} {10}},\ \bibinfo {pages} {261} (\bibinfo {year} {1948})}\BibitemShut
  {NoStop}%
\bibitem [{\citenamefont {Sabisky}\ and\ \citenamefont
  {Anderson}(1973)}]{Sabisky1973}%
  \BibitemOpen
  \bibfield  {author} {\bibinfo {author} {\bibfnamefont {E.~S.}\ \bibnamefont
  {Sabisky}}\ and\ \bibinfo {author} {\bibfnamefont {C.~H.}\ \bibnamefont
  {Anderson}},\ }\bibfield  {title} {\bibinfo {title} {Verification of the
  lifshitz theory of the van der waals potential using liquid-helium films},\
  }\href {https://doi.org/10.1103/PhysRevA.7.790} {\bibfield  {journal}
  {\bibinfo  {journal} {Phys. Rev. A}\ }\textbf {\bibinfo {volume} {7}},\
  \bibinfo {pages} {790} (\bibinfo {year} {1973})}\BibitemShut {NoStop}%
\bibitem [{\citenamefont {Lamoreaux}(1997)}]{Lamoreaux1997}%
  \BibitemOpen
  \bibfield  {author} {\bibinfo {author} {\bibfnamefont {S.~K.}\ \bibnamefont
  {Lamoreaux}},\ }\bibfield  {title} {\bibinfo {title} {Demonstration of the
  casimir force in the 0.6 to $6\ensuremath{\mu}m$ range},\ }\href
  {https://doi.org/10.1103/PhysRevLett.78.5} {\bibfield  {journal} {\bibinfo
  {journal} {Phys. Rev. Lett.}\ }\textbf {\bibinfo {volume} {78}},\ \bibinfo
  {pages} {5} (\bibinfo {year} {1997})}\BibitemShut {NoStop}%
\bibitem [{\citenamefont {Lamoreaux}(1998)}]{LamoreauxE1998}%
  \BibitemOpen
  \bibfield  {author} {\bibinfo {author} {\bibfnamefont {S.~K.}\ \bibnamefont
  {Lamoreaux}},\ }\bibfield  {title} {\bibinfo {title} {Erratum: Demonstration
  of the casimir force in the 0.6 to 6 $\mathit{\ensuremath{\mu}}m$ range
  [phys. rev. lett. 78, 5 (1997)]},\ }\href
  {https://doi.org/10.1103/PhysRevLett.81.5475} {\bibfield  {journal} {\bibinfo
   {journal} {Phys. Rev. Lett.}\ }\textbf {\bibinfo {volume} {81}},\ \bibinfo
  {pages} {5475} (\bibinfo {year} {1998})}\BibitemShut {NoStop}%
\bibitem [{\citenamefont {Lamoreaux}(1999)}]{Lamoreaux1999}%
  \BibitemOpen
  \bibfield  {author} {\bibinfo {author} {\bibfnamefont {S.~K.}\ \bibnamefont
  {Lamoreaux}},\ }\bibfield  {title} {\bibinfo {title} {Calculation of the
  casimir force between imperfectly conducting plates},\ }\href
  {https://doi.org/10.1103/PhysRevA.59.R3149} {\bibfield  {journal} {\bibinfo
  {journal} {Phys. Rev. A}\ }\textbf {\bibinfo {volume} {59}},\ \bibinfo
  {pages} {R3149} (\bibinfo {year} {1999})}\BibitemShut {NoStop}%
\bibitem [{\citenamefont {Spruch}(1993)}]{Spruch1993}%
  \BibitemOpen
  \bibfield  {author} {\bibinfo {author} {\bibfnamefont {L.}~\bibnamefont
  {Spruch}},\ }\bibinfo {title} {An overview of long-range casimir
  interactions},\ in\ \href {https://doi.org/10.1007/978-1-4899-1228-2_1}
  {\emph {\bibinfo {booktitle} {Long-Range Casimir Forces: Theory and Recent
  Experiments on Atomic Systems}}},\ \bibinfo {editor} {edited by\ \bibinfo
  {editor} {\bibfnamefont {F.~S.}\ \bibnamefont {Levin}}\ and\ \bibinfo
  {editor} {\bibfnamefont {D.~A.}\ \bibnamefont {Micha}}}\ (\bibinfo
  {publisher} {Springer US},\ \bibinfo {address} {Boston, MA},\ \bibinfo {year}
  {1993})\ pp.\ \bibinfo {pages} {1--71}\BibitemShut {NoStop}%
\bibitem [{\citenamefont {Milton}(2001)}]{Milton2001}%
  \BibitemOpen
  \bibfield  {author} {\bibinfo {author} {\bibfnamefont {K.~A.}\ \bibnamefont
  {Milton}},\ }\href {https://doi.org/10.1142/4505} {\emph {\bibinfo {title}
  {{The Casimir Effect: Physical Manifestations of Zero-Point Energy}}}}\
  (\bibinfo {year} {2001})\BibitemShut {NoStop}%
\bibitem [{\citenamefont {Moore}(1970)}]{Moore1970}%
  \BibitemOpen
  \bibfield  {author} {\bibinfo {author} {\bibfnamefont {G.~T.}\ \bibnamefont
  {Moore}},\ }\bibfield  {title} {\bibinfo {title} {Quantum theory of the
  electromagnetic field in a variable‐length one‐dimensional cavity},\
  }\href {https://doi.org/10.1063/1.1665432} {\bibfield  {journal} {\bibinfo
  {journal} {J. Math. Phys. (N.Y.)}\ }\textbf {\bibinfo {volume} {11}},\
  \bibinfo {pages} {2679} (\bibinfo {year} {1970})}\BibitemShut {NoStop}%
\bibitem [{\citenamefont {Wilson}\ \emph {et~al.}(2011)\citenamefont {Wilson},
  \citenamefont {Johansson}, \citenamefont {Pourkabirian}, \citenamefont
  {Simoen}, \citenamefont {Johansson}, \citenamefont {Duty}, \citenamefont
  {Nori},\ and\ \citenamefont {Delsing}}]{DynCasEffect2011}%
  \BibitemOpen
  \bibfield  {author} {\bibinfo {author} {\bibfnamefont {C.~M.}\ \bibnamefont
  {Wilson}}, \bibinfo {author} {\bibfnamefont {G.}~\bibnamefont {Johansson}},
  \bibinfo {author} {\bibfnamefont {A.}~\bibnamefont {Pourkabirian}}, \bibinfo
  {author} {\bibfnamefont {M.}~\bibnamefont {Simoen}}, \bibinfo {author}
  {\bibfnamefont {J.~R.}\ \bibnamefont {Johansson}}, \bibinfo {author}
  {\bibfnamefont {T.}~\bibnamefont {Duty}}, \bibinfo {author} {\bibfnamefont
  {F.}~\bibnamefont {Nori}},\ and\ \bibinfo {author} {\bibfnamefont
  {P.}~\bibnamefont {Delsing}},\ }\bibfield  {title} {\bibinfo {title}
  {Observation of the dynamical casimir effect in a superconducting circuit},\
  }\href {https://doi.org/10.1038/nature10561} {\bibfield  {journal} {\bibinfo
  {journal} {Nature}\ }\textbf {\bibinfo {volume} {479}},\ \bibinfo {pages}
  {376} (\bibinfo {year} {2011})}\BibitemShut {NoStop}%
\bibitem [{\citenamefont {Cohen-Tannoudji}\ \emph {et~al.}(1997)\citenamefont
  {Cohen-Tannoudji}, \citenamefont {Dupont-Roc},\ and\ \citenamefont
  {Grynberg}}]{TannoudjiPartOneChapterFour}%
  \BibitemOpen
  \bibfield  {author} {\bibinfo {author} {\bibfnamefont {C.}~\bibnamefont
  {Cohen-Tannoudji}}, \bibinfo {author} {\bibfnamefont {J.}~\bibnamefont
  {Dupont-Roc}},\ and\ \bibinfo {author} {\bibfnamefont {G.}~\bibnamefont
  {Grynberg}},\ }\bibinfo {title} {Other equivalent formulations of
  electrodynamics},\ in\ \href
  {https://doi.org/https://doi.org/10.1002/9783527618422.ch4} {\emph {\bibinfo
  {booktitle} {Photons and Atoms}}}\ (\bibinfo  {publisher} {John Wiley \&
  Sons, Ltd},\ \bibinfo {year} {1997})\ Chap.~\bibinfo {chapter} {4}, pp.\
  \bibinfo {pages} {253--359}\BibitemShut {NoStop}%
\bibitem [{\citenamefont {Barton}(1991)}]{Barton1991}%
  \BibitemOpen
  \bibfield  {author} {\bibinfo {author} {\bibfnamefont {G.}~\bibnamefont
  {Barton}},\ }\bibfield  {title} {\bibinfo {title} {On the fluctuations of the
  casimir force},\ }\href {https://doi.org/10.1088/0305-4470/24/5/014}
  {\bibfield  {journal} {\bibinfo  {journal} {J. Phys. A}\ }\textbf {\bibinfo
  {volume} {24}},\ \bibinfo {pages} {991} (\bibinfo {year} {1991})}\BibitemShut
  {NoStop}%
\bibitem [{\citenamefont {Anglin}\ and\ \citenamefont
  {Zurek}(1996)}]{Anglin1996}%
  \BibitemOpen
  \bibfield  {author} {\bibinfo {author} {\bibfnamefont {J.~R.}\ \bibnamefont
  {Anglin}}\ and\ \bibinfo {author} {\bibfnamefont {W.~H.}\ \bibnamefont
  {Zurek}},\ }\bibfield  {title} {\bibinfo {title} {{A Precision test of
  decoherence}},\ }in\ \href@noop {} {\emph {\bibinfo {booktitle} {{31st
  Rencontres de Moriond: Dark Matter and Cosmology, Quantum Measurements and
  Experimental Gravitation}}}}\ (\bibinfo {year} {1996})\ pp.\ \bibinfo {pages}
  {263--270},\ \Eprint {https://arxiv.org/abs/quant-ph/9611049}
  {arXiv:quant-ph/9611049} \BibitemShut {NoStop}%
\bibitem [{Note1()}]{Note1}%
  \BibitemOpen
  \bibinfo {note} {In the absence of a cutoff, the decoherence kernel takes the
  compact form $\protect \mathcal {D}=\exp {\protect \frac
  {-(d(x')-d(x))^2}{2\pi ^2\hbar \epsilon _0 L^3}\DOTSB \sum@ \slimits@
  _{m=1}^{\infty }\protect \frac {\protect \mathcal {T}}{m^3}\ln (\left
  |\protect \frac {mL+c\protect \mathcal {T}}{mL-c\protect \mathcal {T}}\right
  |)}$. Nevertheless, physically, $\protect \mathcal {D}$ must be obtained by
  setting a finite cutoff as explained in the main text.}\BibitemShut {Stop}%
\bibitem [{\citenamefont {Hornberger}\ \emph {et~al.}(2012)\citenamefont
  {Hornberger}, \citenamefont {Gerlich}, \citenamefont {Haslinger},
  \citenamefont {Nimmrichter},\ and\ \citenamefont
  {Arndt}}]{hornberger2012colloquium}%
  \BibitemOpen
  \bibfield  {author} {\bibinfo {author} {\bibfnamefont {K.}~\bibnamefont
  {Hornberger}}, \bibinfo {author} {\bibfnamefont {S.}~\bibnamefont {Gerlich}},
  \bibinfo {author} {\bibfnamefont {P.}~\bibnamefont {Haslinger}}, \bibinfo
  {author} {\bibfnamefont {S.}~\bibnamefont {Nimmrichter}},\ and\ \bibinfo
  {author} {\bibfnamefont {M.}~\bibnamefont {Arndt}},\ }\bibfield  {title}
  {\bibinfo {title} {Colloquium: Quantum interference of clusters and
  molecules},\ }\href {https://doi.org/10.1103/RevModPhys.84.157} {\bibfield
  {journal} {\bibinfo  {journal} {Rev. Mod. Phys.}\ }\textbf {\bibinfo {volume}
  {84}},\ \bibinfo {pages} {157} (\bibinfo {year} {2012})}\BibitemShut
  {NoStop}%
\bibitem [{\citenamefont {Juffmann}\ \emph {et~al.}(2013)\citenamefont
  {Juffmann}, \citenamefont {Ulbricht},\ and\ \citenamefont
  {Arndt}}]{juffmann2013experimental}%
  \BibitemOpen
  \bibfield  {author} {\bibinfo {author} {\bibfnamefont {T.}~\bibnamefont
  {Juffmann}}, \bibinfo {author} {\bibfnamefont {H.}~\bibnamefont {Ulbricht}},\
  and\ \bibinfo {author} {\bibfnamefont {M.}~\bibnamefont {Arndt}},\ }\bibfield
   {title} {\bibinfo {title} {Experimental methods of molecular matter-wave
  optics},\ }\href {https://doi.org/10.1088/0034-4885/76/8/086402} {\bibfield
  {journal} {\bibinfo  {journal} {Reports on Progress in Physics}\ }\textbf
  {\bibinfo {volume} {76}},\ \bibinfo {pages} {086402} (\bibinfo {year}
  {2013})}\BibitemShut {NoStop}%
\bibitem [{\citenamefont {Gerlich}\ \emph {et~al.}(2007)\citenamefont
  {Gerlich}, \citenamefont {Hackerm{\"u}ller}, \citenamefont {Hornberger},
  \citenamefont {Stibor}, \citenamefont {Ulbricht}, \citenamefont {Gring},
  \citenamefont {Goldfarb}, \citenamefont {Savas}, \citenamefont {M{\"u}ri},
  \citenamefont {Mayor} \emph {et~al.}}]{gerlich2007kapitza}%
  \BibitemOpen
  \bibfield  {author} {\bibinfo {author} {\bibfnamefont {S.}~\bibnamefont
  {Gerlich}}, \bibinfo {author} {\bibfnamefont {L.}~\bibnamefont
  {Hackerm{\"u}ller}}, \bibinfo {author} {\bibfnamefont {K.}~\bibnamefont
  {Hornberger}}, \bibinfo {author} {\bibfnamefont {A.}~\bibnamefont {Stibor}},
  \bibinfo {author} {\bibfnamefont {H.}~\bibnamefont {Ulbricht}}, \bibinfo
  {author} {\bibfnamefont {M.}~\bibnamefont {Gring}}, \bibinfo {author}
  {\bibfnamefont {F.}~\bibnamefont {Goldfarb}}, \bibinfo {author}
  {\bibfnamefont {T.}~\bibnamefont {Savas}}, \bibinfo {author} {\bibfnamefont
  {M.}~\bibnamefont {M{\"u}ri}}, \bibinfo {author} {\bibfnamefont
  {M.}~\bibnamefont {Mayor}}, \emph {et~al.},\ }\bibfield  {title} {\bibinfo
  {title} {A kapitza--dirac--talbot--lau interferometer for highly polarizable
  molecules},\ }\href {https://doi.org/https://doi.org/10.1038/nphys701}
  {\bibfield  {journal} {\bibinfo  {journal} {Nature Physics}\ }\textbf
  {\bibinfo {volume} {3}},\ \bibinfo {pages} {711} (\bibinfo {year}
  {2007})}\BibitemShut {NoStop}%
\bibitem [{\citenamefont {Haslinger}(2013)}]{haslinger2013universal}%
  \BibitemOpen
  \bibfield  {author} {\bibinfo {author} {\bibfnamefont {P.}~\bibnamefont
  {Haslinger}},\ }\bibfield  {title} {\bibinfo {title} {A universal matter-wave
  interferometer with optical gratings}\ }\href
  {https://doi.org/10.25365/thesis.28814} {10.25365/thesis.28814} (\bibinfo
  {year} {2013})\BibitemShut {NoStop}%
\bibitem [{\citenamefont {Kapitza}\ and\ \citenamefont
  {Dirac}(1933)}]{kapitza1933reflection}%
  \BibitemOpen
  \bibfield  {author} {\bibinfo {author} {\bibfnamefont {P.}~\bibnamefont
  {Kapitza}}\ and\ \bibinfo {author} {\bibfnamefont {P.}~\bibnamefont
  {Dirac}},\ }\bibfield  {title} {\bibinfo {title} {The reflection of electrons
  from standing light waves},\ }in\ \href
  {https://doi.org/doi:10.1017/S0305004100011105} {\emph {\bibinfo {booktitle}
  {Mathematical Proceedings of the Cambridge Philosophical Society}}},\
  Vol.~\bibinfo {volume} {29}\ (\bibinfo {organization} {Cambridge University
  Press, Cambridge, England},\ \bibinfo {year} {1933})\ pp.\ \bibinfo {pages}
  {297--300}\BibitemShut {NoStop}%
\bibitem [{\citenamefont {Hornberger}\ \emph {et~al.}(2009)\citenamefont
  {Hornberger}, \citenamefont {Gerlich}, \citenamefont {Ulbricht},
  \citenamefont {Hackermüller}, \citenamefont {Nimmrichter}, \citenamefont
  {Goldt}, \citenamefont {Boltalina},\ and\ \citenamefont
  {Arndt}}]{Hornberger2009}%
  \BibitemOpen
  \bibfield  {author} {\bibinfo {author} {\bibfnamefont {K.}~\bibnamefont
  {Hornberger}}, \bibinfo {author} {\bibfnamefont {S.}~\bibnamefont {Gerlich}},
  \bibinfo {author} {\bibfnamefont {H.}~\bibnamefont {Ulbricht}}, \bibinfo
  {author} {\bibfnamefont {L.}~\bibnamefont {Hackermüller}}, \bibinfo {author}
  {\bibfnamefont {S.}~\bibnamefont {Nimmrichter}}, \bibinfo {author}
  {\bibfnamefont {I.~V.}\ \bibnamefont {Goldt}}, \bibinfo {author}
  {\bibfnamefont {O.}~\bibnamefont {Boltalina}},\ and\ \bibinfo {author}
  {\bibfnamefont {M.}~\bibnamefont {Arndt}},\ }\bibfield  {title} {\bibinfo
  {title} {Theory and experimental verification of
  kapitza–dirac–talbot–lau interferometry},\ }\href
  {https://doi.org/10.1088/1367-2630/11/4/043032} {\bibfield  {journal}
  {\bibinfo  {journal} {New Journal of Physics}\ }\textbf {\bibinfo {volume}
  {11}},\ \bibinfo {pages} {043032} (\bibinfo {year} {2009})}\BibitemShut
  {NoStop}%
\bibitem [{\citenamefont {Hornberger}\ \emph {et~al.}(2003)\citenamefont
  {Hornberger}, \citenamefont {Uttenthaler}, \citenamefont {Brezger},
  \citenamefont {Hackerm\"uller}, \citenamefont {Arndt},\ and\ \citenamefont
  {Zeilinger}}]{ArndtCollDec2003}%
  \BibitemOpen
  \bibfield  {author} {\bibinfo {author} {\bibfnamefont {K.}~\bibnamefont
  {Hornberger}}, \bibinfo {author} {\bibfnamefont {S.}~\bibnamefont
  {Uttenthaler}}, \bibinfo {author} {\bibfnamefont {B.}~\bibnamefont
  {Brezger}}, \bibinfo {author} {\bibfnamefont {L.}~\bibnamefont
  {Hackerm\"uller}}, \bibinfo {author} {\bibfnamefont {M.}~\bibnamefont
  {Arndt}},\ and\ \bibinfo {author} {\bibfnamefont {A.}~\bibnamefont
  {Zeilinger}},\ }\bibfield  {title} {\bibinfo {title} {Collisional decoherence
  observed in matter wave interferometry},\ }\href
  {https://doi.org/10.1103/PhysRevLett.90.160401} {\bibfield  {journal}
  {\bibinfo  {journal} {Phys. Rev. Lett.}\ }\textbf {\bibinfo {volume} {90}},\
  \bibinfo {pages} {160401} (\bibinfo {year} {2003})}\BibitemShut {NoStop}%
\bibitem [{\citenamefont {Hornberger}\ \emph {et~al.}(2004)\citenamefont
  {Hornberger}, \citenamefont {Sipe},\ and\ \citenamefont
  {Arndt}}]{Hornberger2004}%
  \BibitemOpen
  \bibfield  {author} {\bibinfo {author} {\bibfnamefont {K.}~\bibnamefont
  {Hornberger}}, \bibinfo {author} {\bibfnamefont {J.~E.}\ \bibnamefont
  {Sipe}},\ and\ \bibinfo {author} {\bibfnamefont {M.}~\bibnamefont {Arndt}},\
  }\bibfield  {title} {\bibinfo {title} {Theory of decoherence in a matter wave
  talbot-lau interferometer},\ }\href
  {https://doi.org/10.1103/PhysRevA.70.053608} {\bibfield  {journal} {\bibinfo
  {journal} {Phys. Rev. A}\ }\textbf {\bibinfo {volume} {70}},\ \bibinfo
  {pages} {053608} (\bibinfo {year} {2004})}\BibitemShut {NoStop}%
\bibitem [{\citenamefont {Hackerm\"uller}\ \emph {et~al.}(2004)\citenamefont
  {Hackerm\"uller}, \citenamefont {Hornberger}, \citenamefont {Brezger},
  \citenamefont {Zeilinger},\ and\ \citenamefont
  {Arndt}}]{hackermuller2004decoherence}%
  \BibitemOpen
  \bibfield  {author} {\bibinfo {author} {\bibfnamefont {L.}~\bibnamefont
  {Hackerm\"uller}}, \bibinfo {author} {\bibfnamefont {K.}~\bibnamefont
  {Hornberger}}, \bibinfo {author} {\bibfnamefont {B.}~\bibnamefont {Brezger}},
  \bibinfo {author} {\bibfnamefont {A.}~\bibnamefont {Zeilinger}},\ and\
  \bibinfo {author} {\bibfnamefont {M.}~\bibnamefont {Arndt}},\ }\bibfield
  {title} {\bibinfo {title} {Decoherence of matter waves by thermal emission of
  radiation},\ }\href {https://doi.org/10.1038/nature02276} {\bibfield
  {journal} {\bibinfo  {journal} {Nature}\ }\textbf {\bibinfo {volume} {427}},\
  \bibinfo {pages} {711} (\bibinfo {year} {2004})}\BibitemShut {NoStop}%
\bibitem [{\citenamefont {Novotny}\ and\ \citenamefont
  {Hecht}(2012)}]{Novotny_Hecht_2012}%
  \BibitemOpen
  \bibfield  {author} {\bibinfo {author} {\bibfnamefont {L.}~\bibnamefont
  {Novotny}}\ and\ \bibinfo {author} {\bibfnamefont {B.}~\bibnamefont
  {Hecht}},\ }\href@noop {} {\emph {\bibinfo {title} {Principles of
  Nano-Optics}}},\ \bibinfo {edition} {2nd}\ ed.\ (\bibinfo  {publisher}
  {Cambridge University Press, Cambridge, England},\ \bibinfo {year}
  {2012})\BibitemShut {NoStop}%
\bibitem [{\citenamefont {Berninger}\ \emph {et~al.}(2007)\citenamefont
  {Berninger}, \citenamefont {Stefanov}, \citenamefont {Deachapunya},\ and\
  \citenamefont {Arndt}}]{Arndt2007Pol}%
  \BibitemOpen
  \bibfield  {author} {\bibinfo {author} {\bibfnamefont {M.}~\bibnamefont
  {Berninger}}, \bibinfo {author} {\bibfnamefont {A.}~\bibnamefont {Stefanov}},
  \bibinfo {author} {\bibfnamefont {S.}~\bibnamefont {Deachapunya}},\ and\
  \bibinfo {author} {\bibfnamefont {M.}~\bibnamefont {Arndt}},\ }\bibfield
  {title} {\bibinfo {title} {Polarizability measurements of a molecule via a
  near-field matter-wave interferometer},\ }\href
  {https://doi.org/10.1103/PhysRevA.76.013607} {\bibfield  {journal} {\bibinfo
  {journal} {Phys. Rev. A}\ }\textbf {\bibinfo {volume} {76}},\ \bibinfo
  {pages} {013607} (\bibinfo {year} {2007})}\BibitemShut {NoStop}%
\bibitem [{\citenamefont {Israelachvili}(2011)}]{PolarizationBook2011}%
  \BibitemOpen
  \bibinfo {editor} {\bibfnamefont {J.~N.}\ \bibnamefont {Israelachvili}},\
  ed.,\ \href
  {https://doi.org/https://doi.org/10.1016/B978-0-12-391927-4.10024-6} {\emph
  {\bibinfo {title} {Intermolecular and Surface Forces (Third Edition)}}}\
  (\bibinfo  {publisher} {Academic Press},\ \bibinfo {address} {Boston},\
  \bibinfo {year} {2011})\ p.\ \bibinfo {pages} {iii}\BibitemShut {NoStop}%
\bibitem [{\citenamefont {Tikhonov}\ \emph {et~al.}(2001)\citenamefont
  {Tikhonov}, \citenamefont {Kasperovich}, \citenamefont {Wong},\ and\
  \citenamefont {Kresin}}]{Kresin2001}%
  \BibitemOpen
  \bibfield  {author} {\bibinfo {author} {\bibfnamefont {G.}~\bibnamefont
  {Tikhonov}}, \bibinfo {author} {\bibfnamefont {V.}~\bibnamefont
  {Kasperovich}}, \bibinfo {author} {\bibfnamefont {K.}~\bibnamefont {Wong}},\
  and\ \bibinfo {author} {\bibfnamefont {V.~V.}\ \bibnamefont {Kresin}},\
  }\bibfield  {title} {\bibinfo {title} {A measurement of the polarizability of
  sodium clusters},\ }\href {https://doi.org/10.1103/PhysRevA.64.063202}
  {\bibfield  {journal} {\bibinfo  {journal} {Phys. Rev. A}\ }\textbf {\bibinfo
  {volume} {64}},\ \bibinfo {pages} {063202} (\bibinfo {year}
  {2001})}\BibitemShut {NoStop}%
\bibitem [{Mar()}]{MarkusPrivate}%
  \BibitemOpen
  \href@noop {} {\ }\bibinfo {note} {M. Arndt (private
  communications)}\BibitemShut {NoStop}%
\bibitem [{\citenamefont {Marletto}\ and\ \citenamefont
  {Vedral}(2017)}]{MarlettoVedral2017}%
  \BibitemOpen
  \bibfield  {author} {\bibinfo {author} {\bibfnamefont {C.}~\bibnamefont
  {Marletto}}\ and\ \bibinfo {author} {\bibfnamefont {V.}~\bibnamefont
  {Vedral}},\ }\bibfield  {title} {\bibinfo {title} {Gravitationally induced
  entanglement between two massive particles is sufficient evidence of quantum
  effects in gravity},\ }\href {https://doi.org/10.1103/PhysRevLett.119.240402}
  {\bibfield  {journal} {\bibinfo  {journal} {Phys. Rev. Lett.}\ }\textbf
  {\bibinfo {volume} {119}},\ \bibinfo {pages} {240402} (\bibinfo {year}
  {2017})}\BibitemShut {NoStop}%
\bibitem [{\citenamefont {Bose}\ \emph {et~al.}(2017)\citenamefont {Bose},
  \citenamefont {Mazumdar}, \citenamefont {Morley}, \citenamefont {Ulbricht},
  \citenamefont {Toro\ifmmode~\check{s}\else \v{s}\fi{}}, \citenamefont
  {Paternostro}, \citenamefont {Geraci}, \citenamefont {Barker}, \citenamefont
  {Kim},\ and\ \citenamefont {Milburn}}]{Milburn2017}%
  \BibitemOpen
  \bibfield  {author} {\bibinfo {author} {\bibfnamefont {S.}~\bibnamefont
  {Bose}}, \bibinfo {author} {\bibfnamefont {A.}~\bibnamefont {Mazumdar}},
  \bibinfo {author} {\bibfnamefont {G.~W.}\ \bibnamefont {Morley}}, \bibinfo
  {author} {\bibfnamefont {H.}~\bibnamefont {Ulbricht}}, \bibinfo {author}
  {\bibfnamefont {M.}~\bibnamefont {Toro\ifmmode~\check{s}\else \v{s}\fi{}}},
  \bibinfo {author} {\bibfnamefont {M.}~\bibnamefont {Paternostro}}, \bibinfo
  {author} {\bibfnamefont {A.~A.}\ \bibnamefont {Geraci}}, \bibinfo {author}
  {\bibfnamefont {P.~F.}\ \bibnamefont {Barker}}, \bibinfo {author}
  {\bibfnamefont {M.~S.}\ \bibnamefont {Kim}},\ and\ \bibinfo {author}
  {\bibfnamefont {G.}~\bibnamefont {Milburn}},\ }\bibfield  {title} {\bibinfo
  {title} {Spin entanglement witness for quantum gravity},\ }\href
  {https://doi.org/10.1103/PhysRevLett.119.240401} {\bibfield  {journal}
  {\bibinfo  {journal} {Phys. Rev. Lett.}\ }\textbf {\bibinfo {volume} {119}},\
  \bibinfo {pages} {240401} (\bibinfo {year} {2017})}\BibitemShut {NoStop}%
\bibitem [{\citenamefont {Machnikowski}(2006)}]{Machnikowski2006}%
  \BibitemOpen
  \bibfield  {author} {\bibinfo {author} {\bibfnamefont {P.}~\bibnamefont
  {Machnikowski}},\ }\bibfield  {title} {\bibinfo {title} {Theory of which path
  dephasing in single electron interference due to trace in conductive
  environment},\ }\href {https://doi.org/10.1103/PhysRevB.73.155109} {\bibfield
   {journal} {\bibinfo  {journal} {Phys. Rev. B}\ }\textbf {\bibinfo {volume}
  {73}},\ \bibinfo {pages} {155109} (\bibinfo {year} {2006})}\BibitemShut
  {NoStop}%
\bibitem [{\citenamefont {Sonnentag}\ and\ \citenamefont
  {Hasselbach}(2007)}]{Sonnentag2007}%
  \BibitemOpen
  \bibfield  {author} {\bibinfo {author} {\bibfnamefont {P.}~\bibnamefont
  {Sonnentag}}\ and\ \bibinfo {author} {\bibfnamefont {F.}~\bibnamefont
  {Hasselbach}},\ }\bibfield  {title} {\bibinfo {title} {Measurement of
  decoherence of electron waves and visualization of the quantum-classical
  transition},\ }\href {https://doi.org/10.1103/PhysRevLett.98.200402}
  {\bibfield  {journal} {\bibinfo  {journal} {Phys. Rev. Lett.}\ }\textbf
  {\bibinfo {volume} {98}},\ \bibinfo {pages} {200402} (\bibinfo {year}
  {2007})}\BibitemShut {NoStop}%
\bibitem [{\citenamefont {Martinetz}\ \emph {et~al.}(2022)\citenamefont
  {Martinetz}, \citenamefont {Hornberger},\ and\ \citenamefont
  {Stickler}}]{Lukas2022}%
  \BibitemOpen
  \bibfield  {author} {\bibinfo {author} {\bibfnamefont {L.}~\bibnamefont
  {Martinetz}}, \bibinfo {author} {\bibfnamefont {K.}~\bibnamefont
  {Hornberger}},\ and\ \bibinfo {author} {\bibfnamefont {B.~A.}\ \bibnamefont
  {Stickler}},\ }\bibfield  {title} {\bibinfo {title} {Surface-induced
  decoherence and heating of charged particles},\ }\href
  {https://doi.org/10.1103/PRXQuantum.3.030327} {\bibfield  {journal} {\bibinfo
   {journal} {PRX Quantum}\ }\textbf {\bibinfo {volume} {3}},\ \bibinfo {pages}
  {030327} (\bibinfo {year} {2022})}\BibitemShut {NoStop}%
\bibitem [{\citenamefont {Jakubec}\ \emph {et~al.}(2024)\citenamefont
  {Jakubec}, \citenamefont {Jarzynski},\ and\ \citenamefont
  {Sinha}}]{Jakubec2024}%
  \BibitemOpen
  \bibfield  {author} {\bibinfo {author} {\bibfnamefont {C.}~\bibnamefont
  {Jakubec}}, \bibinfo {author} {\bibfnamefont {C.}~\bibnamefont {Jarzynski}},\
  and\ \bibinfo {author} {\bibfnamefont {K.}~\bibnamefont {Sinha}},\ }\bibfield
   {title} {\bibinfo {title} {{Decoherence and Brownian motion of a polarizable
  particle near a surface}},\ }\href@noop {} {\  (\bibinfo {year} {2024})},\
  \Eprint {https://arxiv.org/abs/2408.15433} {arXiv:2408.15433 [quant-ph]}
  \BibitemShut {NoStop}%
\end{thebibliography}%
\newpage
\begin{widetext}
\begin{center}
\textbf{End Matter}
\end{center}
\end{widetext}
\textit{Appendix A: Computation of the decoherence kernel $\mathcal{D}$}.-- For a polarizable particle located at coordinate $x$ near the center ($|x|\ll L$), with a time dependent dipole moment along the x-axis $d^{x}(x,t)$, the state of the EM field at time $t$, to leading order in the interaction picture, under the dipole approximation, is given by  
\begin{align}\label{App:EnvState}
\ket{\mathcal{E}(x)}_{t} \approx \exp(\frac{i}{\hbar}\int_{0}^{t} dt'd^{x}(x,t')\hat{\Pi}^{x}(0,t') )\ket{0}.
\end{align}
The time dependence of the operator $\hat{\mathbf{\Pi}}$ comes only through $e^{-i\omega_n t}$, while that of the dipole moment comes via the switching function $s(t)$ specified in Eq.~\eqref{AEQ:Switch} of the main text. Therefore, to compute $\mathcal{D}(x,x',t)$, we must evaluate the sum $\sum_{m=1}^{N}$ over
\begin{align}\label{AEQ:Time_Int}
(-1)^m\int_{0}^{t} dt' e^{-i\omega_n t'} \theta(t'-t_m)=(-1)^m\int_{t_m}^{t}dt' e^{-i\omega_n t'}.    
\end{align} 
The integer $N$ is even, since at the end of the process we must have $d^x(x,t>t_N)=0$. Therefore, for $t>t_N$,  
\begin{align}\label{supp:TimeIntEnvState}
\sum_{m=1}^{N}(-1)^m\int_{t_m}^{t} dt'e^{-i\omega_n t'}  = \frac{e^{-i\phi}}{\omega_n}\frac{\sin{(N \omega_n\mathcal{T}/2)}}{\cos{(\omega_n\mathcal{T}/2)}},
\end{align} 
where $\phi:=\pi+(N+1)\omega_n\mathcal{T}/2$. Having computed the time integral in Eq.~\eqref{AEQ:TimeEv} and/or Eq.~\eqref{App:EnvState}, we now calculate the inner product in Eq.~\eqref{AEQ:DecKernel}. 

As specified in Eq.~\eqref{AEQ:ElecField}, $\hat{\Pi}^{x}$ is a linear sum of creation and annihilation operators. Thus, in the Fock space, the state of each of the vacuum modes is analogous to the coherent state 
\begin{align}
\ket{\alpha_{\k_{\parallel}n}(x,t)} &= \exp{\alpha_{\k_{\parallel}n}(x,t)\hat{a}^{\dagger}_{2}(\k_{\parallel},n)-\mathrm{c.c.}}\ket{0}.
\end{align}
For a given mode characterized by the integer $n$ and the wave-vector $\k_{\parallel}$, using the result in Eq.~\eqref{supp:TimeIntEnvState} for the time integral in Eq.~\eqref{App:EnvState}, we get
\begin{align}\label{app:CoherentState}
\alpha_{\k_{\parallel}n}(x,t) = \frac{d(x)f(n)k_{\parallel}c\cos\left(\frac{n\pi}{2}\right)}{2\pi\sqrt{\omega^3_n\hbar\epsilon_0 L}}\frac{\sin\left(\frac{N\omega_n\mathcal{T}}{2}\right)e^{-i\phi}}{\cos\left(\frac{\omega_n\mathcal{T}}{2}\right)}.
\end{align}
Using the standard formula for the inner product between coherent states $\ket{\alpha_1}$ and $\ket{\alpha_2}$,
\begin{align}
\bra{\alpha_1}\ket{\alpha_2} = e^{-(|\alpha_1|^2+|\alpha_2|^2-2\alpha_1^*\alpha_2)/2},
\end{align}
we get for the decoherence kernel
\begin{align}\label{app:OverlapAdiabatic}
\mathcal{D}&=\bra{\mathcal{E}(x')}\ket{\mathcal{E}(x)}_{t>t_N} =\prod_{\k_{\parallel}}\prod_{n}\bra{\alpha_{\k_{\parallel} n}(x',t)}\ket{\alpha_{\k_{\parallel} n}(x,t)}\nonumber\\
&=\exp\left\lbrace\frac{-(d_{x'}-d_{x})^2}{4\pi^2\hbar \epsilon_0}\sum_{n=0}^{\infty}\int d\k_{\parallel}f^2(n)\mathcal{I}\left(|k_{\parallel}|,|n|\right)\right\rbrace,
\end{align}
where $d_x$ stands for $d_x := d(x)$, and 
\begin{align}
\mathcal{I}\left(|k_{\parallel}|,|n|\right) := \frac{(k_{\parallel }c)^2\cos^2(n\pi/2)}{ 2L\omega^3_{n}}\frac{\sin^2(N\omega_n\mathcal{T}/2)}{\cos^2(\omega_n\mathcal{T}/2)}.
\end{align}
Further, since $\mathcal{I}\left(|k_{\parallel}|,|n|\right)$ is an even function of the integer $n$, the sum over the integers appearing in the equation above can be written as
\begin{align}\label{Supp:WeightedInt}
&\sum_{n=0}^{\infty}f^2(n)\mathcal{I}\left(|k_{\parallel}|,|n|\right) =\nonumber\\ &\frac{1}{2}\sum_{n=-\infty}^{-1}\mathcal{I}\left(|k_{\parallel}|,|n|\right)+\frac{1}{2}\mathcal{I}\left(|k_{\parallel}|,0\right) +\frac{1}{2}\sum_{n=1}^{\infty}\mathcal{I}\left(|k_{\parallel}|,|n|\right).
\end{align}
The $n=0$ term gets an additional weight of $1/2$ due to the presence of $f^2(n)$ in the sum. Using Eq.~\eqref{Supp:WeightedInt}, the decoherence kernel can be written as in Eq.~\eqref{AEQ:OverlapAdiabatic} of the main text. To evaluate the integral, it is convenient to convert the surface integral in $\k_{\parallel}$ into the standard volume integral in $\k$, by writing $\int d\k_{\parallel}$ as $\int d\k\delta(k_{x}-n\pi/L)$ and $\omega^2_n/c^2 = \k^2_{\parallel}+n^2\pi^2/L^2$ as $ \k\cdot\k=k^2$. Doing so, the integral in Eq.~\eqref{AEQ:OverlapAdiabatic} becomes
\begin{align}
&\frac{1}{L}\sum_{n=-\infty}^{\infty}\int d\k_{\parallel}\frac{(k_{\parallel }c)^2\cos^2(n\pi/2)}{ 4\omega^3_{n}}\frac{\sin^2(N\omega_n\mathcal{T}/2)}{\cos^2(\omega_n\mathcal{T}/2)} =\nonumber\\
&\frac{1}{L}\int d\k\frac{(k^2-k^2_x)c^2}{ 4k^3c^3}\frac{\sin^2(Nkc\mathcal{T}/2)}{\cos^2(k c\mathcal{T}/2)}\sum_{n=-\infty}^{\infty}\cos^2(n\pi/2)\delta_n,
\end{align}
where $\delta_n:=\delta(k_x-n\pi/L)$, as defined in Eq.~\eqref{AEQ:DiracComb}. Using the identity in Eq.~\eqref{AEQ:DiracComb}, and setting $N=2$, the integral above can be written in spherical coordinates as 
\begin{align}\label{app:MainInt}
&\frac{1}{L}\int d\k\frac{(k^2-k^2_x)c^2}{ 4k^3c^3}\frac{\sin^2(Nkc\mathcal{T}/2)}{\cos^2(k c\mathcal{T}/2)}\sum_{n=-\infty}^{\infty}\cos^2(n\pi/2)\delta_n\nonumber\\
&=\frac{1}{c}\sum_{m=-\infty}^{\infty}\int_{0}^{k_{\mathrm{max}}} dk k\sin^2(kc\mathcal{T}/2) \int_{-1}^{1} du(1-u^2)e^{imkLu}.
\end{align}
The $\cos^2(k c\mathcal{T}/2)$ which was previously present in the denominator goes away after setting $N=2$ and using the identity $\sin(2\theta) = 2\sin(\theta)\cos(\theta)$. The main result, in Eq.~\eqref{AEQ:DecCutoff} of the main text, is obtained by imposing the cutoff $k_{\text{max}}$ in the radial part of the integral and discarding the $m=0$ term. As mentioned in the main text, the $m=0$ term is independent of the plate separation $L$, and would correspond to false decoherence in empty space. As shown in \cite{Gundhi:2023vjs}, working within the dipole approximation,  unless the interaction is switched on and off for this term on a time scale comparable to or much less than $1/(k_\text{max}c)$, this term corresponds to false decoherence. The non-adiabatic switching on-off for the $m=0$ term is not achieved, since the time scale over which the interaction is physically switched on and off in the setup is given by $a/v_{z}\simeq 10^{-11}$s, which is much greater than $1/(k_\text{max}c)\simeq 10^{-18}$s. 

\textit{Appendix B: Decoherence kernel without the dipole approximation}.-- In this section our aim is to show that in certain special cases, the analytic expression for $\mathcal{D}$
can be obtained without resorting to the dipole approximation. The special case we present here is when superpositions are prepared near the plates such that
\begin{align}
&\ket{\mathcal{E}(\pm L/2)}_{t} =\nonumber\\ &\exp(\frac{i}{\hbar}\int_{0}^{t} dt'd^{x}(\pm L/2,t')\hat{\Pi}^{x}(\pm L/2,t') )\ket{0}.
\end{align}
In this case,
\begin{align}\label{app:CoherentStateND}
&\alpha_{\k_{\parallel}n}(-L/2,t) = \frac{d(-L/2)f(n)k_{\parallel}c}{2\pi\sqrt{\omega^3_n\hbar\epsilon_0 L}}\frac{\sin\left(\frac{N\omega_n\mathcal{T}}{2}\right)e^{-i\phi}}{\cos\left(\frac{\omega_n\mathcal{T}}{2}\right)},\nonumber\\
&\alpha_{\k_{\parallel}n}(L/2,t) = (-1)^{n}\frac{d(L/2)f(n)k_{\parallel}c}{2\pi\sqrt{\omega^3_n\hbar\epsilon_0 L}}\frac{\sin\left(\frac{N\omega_n\mathcal{T}}{2}\right)e^{-i\phi}}{\cos\left(\frac{\omega_n\mathcal{T}}{2}\right)}.
\end{align}
Following the same procedure as before, similar to Eq.~\eqref{app:MainInt}, the decoherence kernel is obtained by evaluating the integral (keeping the additional factor analogous to $(d_{x'}-d_{x})^2/2$)
\begin{align}
&\frac{d^2_{\frac{L}{2}}+d^2_{\frac{-L}{2}}}{2L}\int d\k\frac{(k^2-k^2_x)c^2}{ 4k^3c^3}\frac{\sin^2(Nkc\mathcal{T}/2)}{\cos^2(k c\mathcal{T}/2)}\sum_{n=-\infty}^{\infty}\delta_n-\nonumber\\
&\frac{d_{\frac{L}{2}}d_{\frac{-L}{2}}}{L}\int d\k\frac{(k^2-k^2_x)c^2}{ 4k^3c^3}\frac{\sin^2(Nkc\mathcal{T}/2)}{\cos^2(k c\mathcal{T}/2)}\sum_{n=-\infty}^{\infty}(-1)^n\delta_n.
\end{align}
However, if we now assume that the profile of the laser light is such that $d_{\frac{-L}{2}}=-d_{\frac{L}{2}}$, for instance if the electric field of the laser light $\propto \sin(\pi x/l)$ -- which is typical for laser gratings and is not a restrictive assumption -- then the integral above becomes 
\begin{align}
\frac{|d_{\frac{L}{2}}-d_{\frac{-L}{2}}|^2}{2L}\int d\k\frac{(k^2-k^2_x)c^2}{ 4k^3c^3}\frac{\sin^2(Nkc\mathcal{T}/2)}{\cos^2(k c\mathcal{T}/2)}\sum_{n\text{ even}}\delta_n,
\end{align}
which is the same as in Eq.~\eqref{app:MainInt}. This estimate of decoherence, which retains the spatial dependence of the EM field, shows that the dipole approximation does not necessarily lead to drastic differences in the expression for $\mathcal{D}$. In fact, for the special case studied here, $\mathcal{D}$ is obtained by simply replacing $(d(x')-d(x))^2$ by $(d(-L/2)-d(L/2))^2$ in Eq.~\eqref{AEQ:DecCutoff}. Even though this scenario is not optimal, since unwanted effects due to images might become dominant close to the conducting plates at $x=\pm L/2$, it shows that the functional form of the decoherence kernel is not necessarily strongly dependent on  the application of the dipole approximation.

 \textit{Appendix C: Competing effects; estimate for decoherence due to image charges}.--  To the list of standard competing decoherence effects \cite{ArndtCollDec2003} which are encountered and typically overcome in matter-wave interferometry, the setup that we propose only adds image currents \cite{Anglin1996,Machnikowski2006,Sonnentag2007,Lukas2022,Jakubec2024}.
A charged particle between conducting plates leads to an infinite number of image charges as per standard electrostatics. The images (or equivalently the free charges inside the metal) must instantly move with the charged particle in order to nullify its Coulomb field along the plates. The free charges inside the conductor are therefore correlated only to the position of the particle and not its trajectory. They cannot acquire which-path information and lead to no observable decoherence themselves \cite{Gundhi2024Casimir}. Coherence can still be lost, however, if the plates do not respect the ideal conductor assumption, and offer a finite resistance to the image currents. In such a scenario, the charged particle heats the region of the conductor that it passes by, due to the friction encountered by the induced image charges. This enables the conductor to \textit{record} the particle's which-path information, thus leading  to observable decoherence \cite{Anglin1996,Machnikowski2006,Sonnentag2007,Lukas2022,Jakubec2024}.
However, decoherence due to images decreases as the distance of the charged particle from the plates increases. As we show next, decoherence due to image currents will be negligible in our proposal.

In \cite{Anglin1996}, decoherence due to image currents, induced by the motion of a point charged particle,  was estimated. The scenario considered in this work, instead, involves motion of dipole moments. Connection to the estimates made in \cite{Anglin1996} can be made by noticing that the electric field due to a dipole, at large distances $R$, is proportional to $|\mathbf{d}|/R^3$, while that of a point charge $Q$ is proportional to $Q/R^2$. Thus, the magnitude of the induced image charge,  in order to nullify the electric field along the conducting plates, can be estimated to be around $Q\simeq |\mathbf{d}|/L$. This is because for superpositions near the center, the distance of the dipole from the plates is $R\simeq L$. For alkali earth clusters where $|\mathbf{d}|\simeq 10^{-22}$ C m, we get $Q\simeq 10^{-22}/10^{-3}$C $\simeq$ $1$e, for $L\simeq 10^{-3}$m.

The formula for the decoherence time scale for a single ion, $\tau_{d} = \left(10^{4} x/\text{m}\right)^3\times 10^{-5}$ s, is obtained in \cite{Anglin1996} which corresponds to $\tau_{d}\simeq 0.01$ s for $x\simeq 10^{-3}$ \text{m}. This time scale is much longer than the time $\mathcal{T} \simeq \sigma_z/v_{z}= 10^{-9}$ s, for which the dipole moment of the molecules would be switched on in our proposal. Since the time of flight for the dipole $\mathcal{T}$  is much less than the decoherence time $\tau_d$, we expect decoherence due to VFs to dominate, and not be suppressed by the images. 

\end{document}